\documentstyle[osa,manuscript]{revtex}
\input epsf
\begin{document}

\title{A waveguide atom beamsplitter for laser-cooled neutral atoms}

\author{\renewcommand{\thefootnote}{\fnsymbol{footnote}}
Dirk M\"{u}ller,  Eric A. Cornell,\footnote[1]{Also at Quantum
Physics Division, National Institute of Standards and Technology,
Boulder, CO 80309} Marco Prevedelli,\footnote[2]{Permanent
address: Dipartimento di Fisica, Dell' Universit\`a di Firenze,
Largo E. Fermi 2, 50125 Firenze, Italy} Peter D. D. Schwindt, Alex
Zozulya,\footnote[3]{Present address: Department of Physics,
Worchester Polytechnic Institute, Worchester, MA 01609} Dana Z.
Anderson}
\renewcommand{\thefootnote}{\arabic{footnote}}

\address{Department of Physics, and JILA, University of Colorado and NIST,
Boulder, CO 80309-0440}

\date{\today}

\maketitle
\begin{abstract}

A laser-cooled neutral-atom beam from a low-velocity intense
source is split into two beams while guided by a magnetic-field
potential.  We generate our multimode-beamsplitter potential with
two current-carrying wires on a glass substrate combined with an
external transverse bias field. The atoms bend around several
curves over a $10$-cm distance. A maximum integrated flux of
$1.5\cdot10^{5} \mathrm{atoms/s}$ is achieved with a current
density of $5\cdot10^{4} \mathrm{Ampere/cm^{2}}$ in the
100-$\mathrm{\mu m}$ diameter wires.  The initial beam can be
split into two beams with a 50/50 splitting ratio.

\end{abstract}

\narrowtext

Like their optical counterpart, atom beamsplitters are the pivotal
element of atom-optical interferometers.  While the original
beamsplitter was perhaps the Stern-Gerlach apparatus
\cite{Stern-Gerlach}, modern free-space beamsplitters are based on
mechanical or light-based refractive elements. Such beamsplitters
have been used with good success in Mach-Zehnder interferometers
to measure the Sagnac-effect with high sensitivity
\cite{Pritchard,Kasevich}. Free-space beamsplitters are generally
characterized by small splitting angles because the effective
grating spacing is large compared to the atomic de Broglie
wavelength.

A waveguide-based beamsplitter has the potential to provide
arbitrary splitting angles.  Furthermore, the confining potential
of a waveguide also suppresses the beam divergence and
gravitational sag to which free-space interferometers are
subjected.  Several atom-guiding schemes using magnetic forces
have been proposed and demonstrated
\cite{Schmiedmayer,Hinds,DenschlagApplPhys,DenschlagPRL,Prentiss,Haensch1,Haensch2}.
We recently demonstrated guiding a beam of laser-cooled atoms
around a curve \cite{Mueller2} using magnetic forces from
photolithographically patterned current-carrying wires.  The
multimode-atom beamsplitter reported on here is a natural
extension of our previously demonstrated guiding scheme.  Like its
fiber and integrated optical counterparts, our waveguide
beamsplitter merges and then diverges two guiding regions.

We guide $^{87}$Rb atoms in a weak-field-seeking state along a
magnetic-field minimum. This magnetic guide leads atoms around
several curves to a beamsplitter region.  Our beamsplitter region
consists of two such magnetic-field minima that merge to one field
minimum and separate again into two minima. A variable fraction of
atoms initially launched into one of the two magnetic field minima
are guided and transferred into the second magnetic field minimum
at the beamsplitter region.

For our atom source we prepare a laser-cooled beam of
$^{87}\rm{Rb}$ atoms with a modified vapor-cell magneto-optical
trap (MOT) \cite{MOT}.  To generate our low-velocity intense
source (LVIS)\cite{LVIS} we drill a 500-$\rm{\mu m}$ hole in the
center of a retro-reflecting mirror placed inside our vacuum
chamber [Fig. 1(d)]. We couple LVIS atoms into our magnetic guide
by positioning the guide opening directly behind the mirror hole.
The atom's internal state and velocity distributions are as
measured previously \cite{Mueller2,Mueller1}.

We generate our one-dimensional guiding potential by adding an
external transverse bias field to the magnetic field generated by
a 100$\times$100-micron current-carrying wire on a glass substrate
\cite{DenschlagPRL,Prentiss,Haensch2}. The vector sum of the
transverse bias field $\vec{B}_{bias}$ and the wire's magnetic
field $\vec{B}_{wire}$ becomes zero at a position outside the
wire, if the bias field is smaller than the field generated by the
wire at its surface [Fig. 1(a)]. As the bias field is increased
(decreased) the position of the magnetic field minimum moves
linearly toward (away from) the wire and the potential depth
increases (decreases). Furthermore, when the wire's magnetic-field
maximum is twice the transverse bias field, the magnetic field
zero is 50-$\rm{\mu m}$ above the current-carrying-wire surface
and the field magnitude increases linearly with displacement in
the transverse directions.  We generate the transverse bias field
for the guide with an electromagnet placed near the substrate
[Fig. 1(d)].  An additional $\sim$14-G longitudinal bias field is
applied to prevent the magnetic-field magnitude from vanishing at
the field minimum.  As the wire current and bias field are
increased proportionally the magnetic-potential depth and gradient
increase linearly, but the field-minimum position remains
unchanged.

With current only in the wire positioned right behind the mirror
hole, wire 1, and a 86-G external bias field applied we guide
atoms and measure atom flux versus wire current for different bias
fields. In this guiding experiment we run 35-msec-long current
pulses of up to 5.5 A at a 1 sec repetition rate through wire 1.
We choose short current pulses to prevent the glass substrate from
overheating, allowing us to run larger wire currents than a
continuous current would allow. After the atoms exit the guide,
they are ionized by a hot wire and the subsequent ions are then
detected by a channeltron.  For each external bias-field value
there is an optimum track current that maximizes the guided-atom
flux [Fig. 2(a)]. For wire currents too large the magnetic-field
minimum is shifted far away from the wire resulting in a reduced
field gradient.  This field-gradient reduction helps to couple
atoms into the guide opening, but also leads to guiding losses as
atoms can no longer be guided around the curves of the guide. When
the track current is too low the generated magnetic-field gradient
is sufficient to bend the atoms around the curve, but the field
minimum is close to the wire surface and atom-surface interactions
as well as a tighter guide opening result in a lower flux.  Our
guiding flux peaks when the condition for mode matching atoms into
the guide opening and maintaining a sufficient gradient to bend
atoms around the curves is optimized.

We measure the heating of guided atoms by comparing the transverse
velocities before and after the guiding process. At a wire current
of 5.0 A and a transverse bias field of $\sim86$ G we measure the
guided-atoms' transverse-velocity profile by translating the hot
wire transverse to the propagation direction to map out the
spatial extent of the atom beam as it diverges from the guide exit
[Fig. 2(b)]. The 70-$\mathrm{\mu m}$-diameter hot wire is placed
$\sim2.5$ cm from the output of the magnetic guide.  We calculate
that the atoms' emergence from the confining fields of the guide
is almost completely non-adiabatic---the transverse kinetic energy
of the emerging beam should thus be a faithful reflection of the
transverse kinetic energy in the guide. From the width of the fit
in Figure 2(b) we determine that the transverse-velocity
distribution of the guided atoms is $v_t=17.2\pm
3.5\,\,\mathrm{cm/sec}$, in contrast to an initial transverse
velocity of $v_t=10.0\pm 1.5\,\,\mathrm{cm/s}$ of LVIS.  We
attribute the observed heating to the non-adiabatic loading of the
LVIS atoms into the guide.  An atom that enters the guide
displaced from the magnetic field minimum experiences a sudden
increase in its potential energy because it is not mode-matched to
the guide. The additional potential energy increases its total
energy inside the guide, which is converted into transverse
kinetic energy once the atom leaves the guide. We believe this
heating effect can be ameliorated by adiabatically loading the
atoms into a tapered guide.

Once atoms are guided along the one-dimensional magnetic-field
minimum we turn on our beamsplitter.  As we increase the current
running through wire 2 we observe that the flux from guide 2
increases and the flux out of guide 1 decreases. Figure 3(a) shows
the flux of guided atoms coming out of each guide versus the
current ratio between the two wires. As we change this current
ratio we can tune the splitting ratio of our beamsplitter.  We
observe a dynamic range of the splitting ratio from 100/0 to
15/85. A 50/50 beamsplitter is achieved when the current in wire 2
is 85\% of the current in wire 1.  Aside from the current ratio
there are two other parameters that determine the splitting ratio
of our beamsplitter. First, as we change the applied transverse
bias field we can vary the degree of overlap of the two magnetic
field minima. For very large bias fields the two minima remain
close to their respective wires and their overlap is small in the
beamsplitter region. Second, the curvature of the guides in the
beamsplitter region determines the manner in which the two
magnetic-field minima merge.  In our design the atoms are
preferentially switched over into the secondary guide due to the
wire curvature in the beamsplitter region.  The guides bend with a
radius of curvature of $\sim$30 cm into the splitter region and
curve away with a radius of $\sim$70 cm. A calculation of our
potential-minima trajectories shows that for this curvature the
field minima follow straight lines that cross (Fig. 4). This means
that atoms launched into guide 1, the primary guide of our
beamsplitter, are more likely to switch over into guide 2, the
secondary guide, than to continue along guide 1. This geometric
feature of our beamsplitter is responsible for its bias toward
coupling atoms into guide 2.  Our experimental data shows that at
a current ratio of 0.85 between guide 2 and guide 1 we compensate
for the geometric bias of our beamsplitter and achieve 50/50
beamsplitting.  As we increase the current in wire 2 to a current
larger than in wire 1 the flux out of guide 2 peaks at a current
ratio $I_{wire 2}/I_{wire 1}=1.1$ and decreases again beyond this
value. This decrease is analogous to the observation in 2(a) where
we found a decrease in guided-atom flux for wire currents in a
non-optimum $B_{wire}/B_{bias}$-ratio regime.

We sum the flux from the output of both guides and find the total
flux to remain roughly constant.  This observed flux conservation
of our beamsplitter as shown in Figure 3(a) is a strong function
of the transverse bias field applied. For a current in wire 1 of
5.0 A and a longitudinal bias field of 14 G a transverse bias
field of $\sim86$ G is necessary to achieve a constant total flux
as the current ratio is varied. This is 35\% larger than the
optimum bias field measured for maximum guiding efficiency when
only one wire carries current.  We postulate that the increased
bias field separates the two magnetic field minima in the
beamsplitter region making it easier for atoms to follow guide 1
instead of guide 2.

We model our beamsplitter design with a numerical simulation and
find qualitative agreement with our measured data.  The simulation
calculates the classical trajectories of $2.5 \times 10^6$ atoms
through our beamsplitter potential and records which guide each
atom exits [Fig. 3(b)]. The test atoms' spatial initial conditions
are spread over a $100 \times 100$-$\mu \mathrm{m}$ aperture
centered 60 $\mu \mathrm{m}$ above the wire and their initial
velocities match the LVIS-velocity distribution at the guide
opening. We use constant bias fields of 100 G in the transverse
and 14 G in the longitudinal direction for the simulation shown.
In our experiment these values are not constant along the entire
guiding region, which limits the accuracy of our simulation.  The
simulation shows that more atoms are coupled into guide 2 than
remain in guide 1 when the currents in both wires are equal, which
confirms our idea that the beamsplitter is biased toward guide 2
by its geometric design. Further, a 50/50 beamsplitting is shown
at a current ratio of 0.75, which compares well within the errors
of the simulation to the measured current ratio of 0.85.  Our
simulation also shows a decrease in flux out of both guides at the
largest current ratios as observed in our experiment.

In summary, we have split a beam of laser-cooled atoms while
guided in a magnetic-field potential.  We are able to vary the
ratio of atoms coupled into the secondary guide from 0 to
$\sim$85$\%$ by varying the current in wire 2.  We demonstrate
that moderate currents give guiding potential depths of several
millidegrees Kelvin.  In the future the question of coherent
beamsplitting will have to be addressed. While coherence has been
well established for free-space atom interferometers it has yet to
be demonstrated that our waveguide beamsplitter preserves
coherence.

\smallskip

The authors would like to thank Carl Wieman and Randal Grow for
helpful discussions.  This work was made possible by funding from
the Office of Naval Research (Grant No. N00014-94-1-0375) and the
National Science Foundation (Grant No. Phy-95-12150).

\pagebreak

\begin{figure}
\epsfxsize=3.5 truein \epsfbox{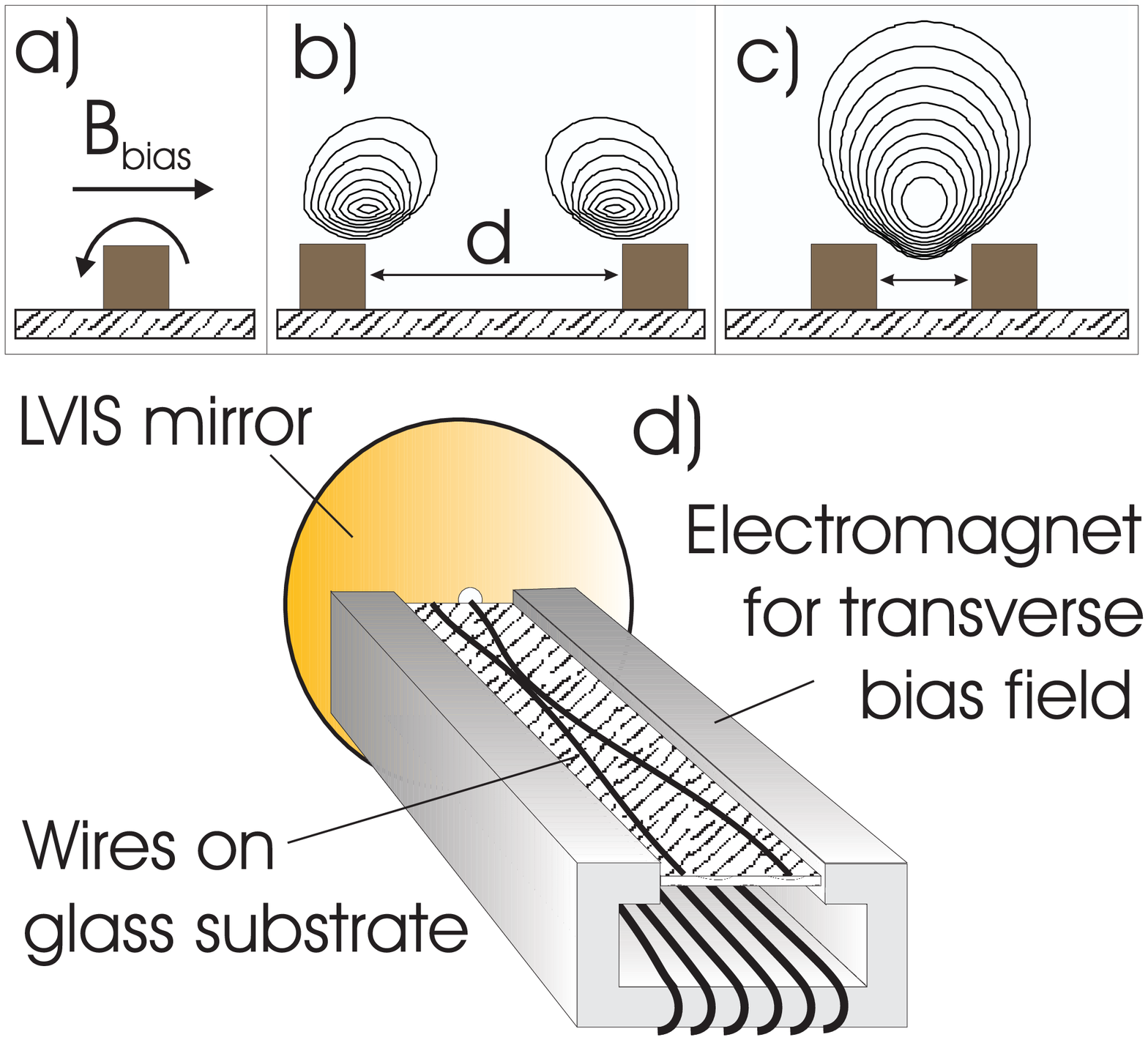}

\caption{Contour lines of magnetic-field potential and guide
schematic. (a-c) We show a cross-sectional cut across the wires.
When a bias field is applied transverse to the wires the magnetic
field becomes zero just above the wire surface (a).  For large
track separations (d=300$\mu \rm{m}$) the magnetic field  minima
do not merge (b). As the track spacing is reduced (d=100$\mu
\rm{m}$) the magnetic field minima merge to form one field minimum
(c).  The transverse bias field is generated with an electromagnet
near the wire substrate (d).  The LVIS mirror hole is aligned with
one of the wires to couple the LVIS atoms into the guide.
\label{fig1}}
\end{figure}

\pagebreak

\begin{figure}
\epsfxsize=3.8 truein \epsfbox{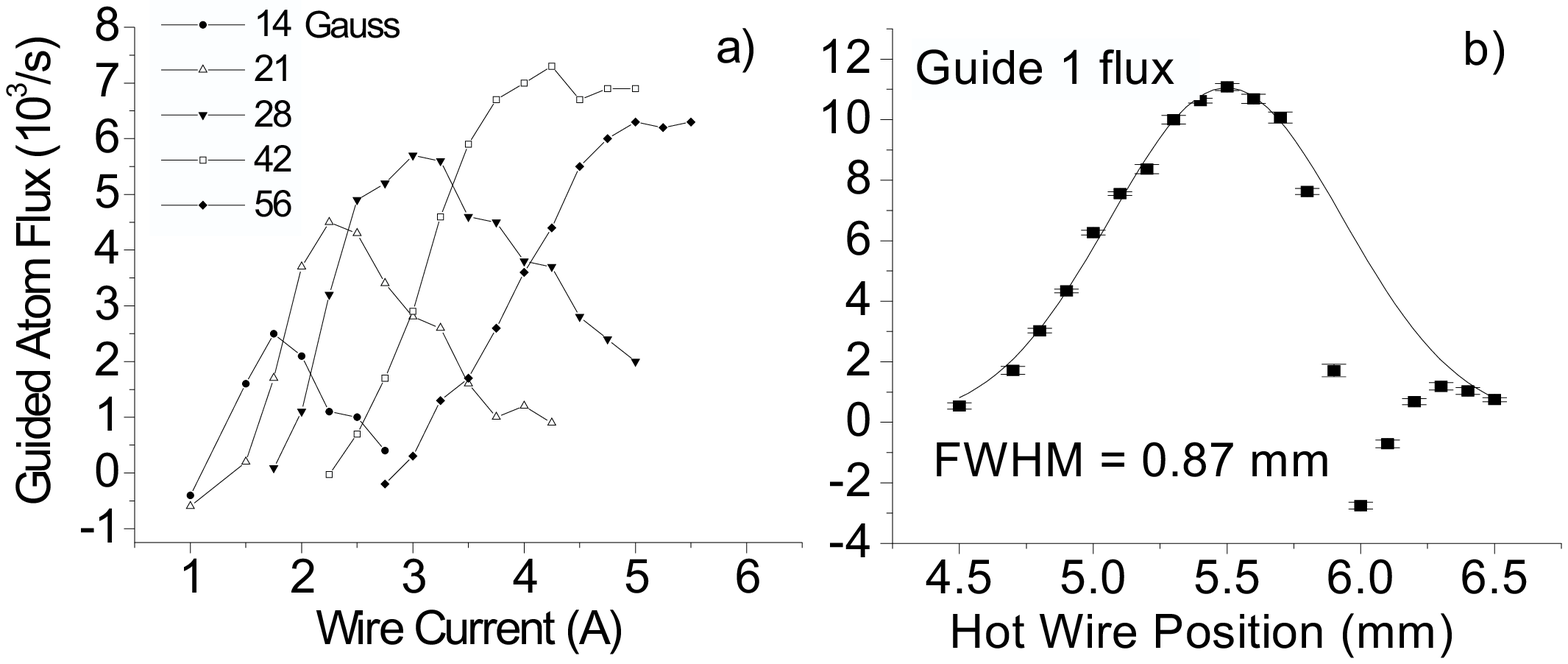}

\caption{Guided atom flux. We measure the guided atom flux versus
the wire current at different transverse bias fields.  For each
bias field we observe an optimum wire current where the guided
flux peaks (a). Figure 2(b) shows the guided-atom beam profile. We
move the hot wire perpendicular to the propagation direction to
map out the transverse-velocity distribution.  We use the width of
the Gaussian fit to determine the RMS transverse velocity to be
$v_t=17.2\pm 3.5 \rm{cm/sec}$. The dip in flux around 6 mm is an
artifact due to some stray LVIS atoms that contaminate the
guided-atom-beam profile.  Those data points appear negative due
to a background-subtraction procedure. \label{fig2}}
\end{figure}

\pagebreak

\begin{figure}
\epsfxsize=3.5 truein \epsfbox{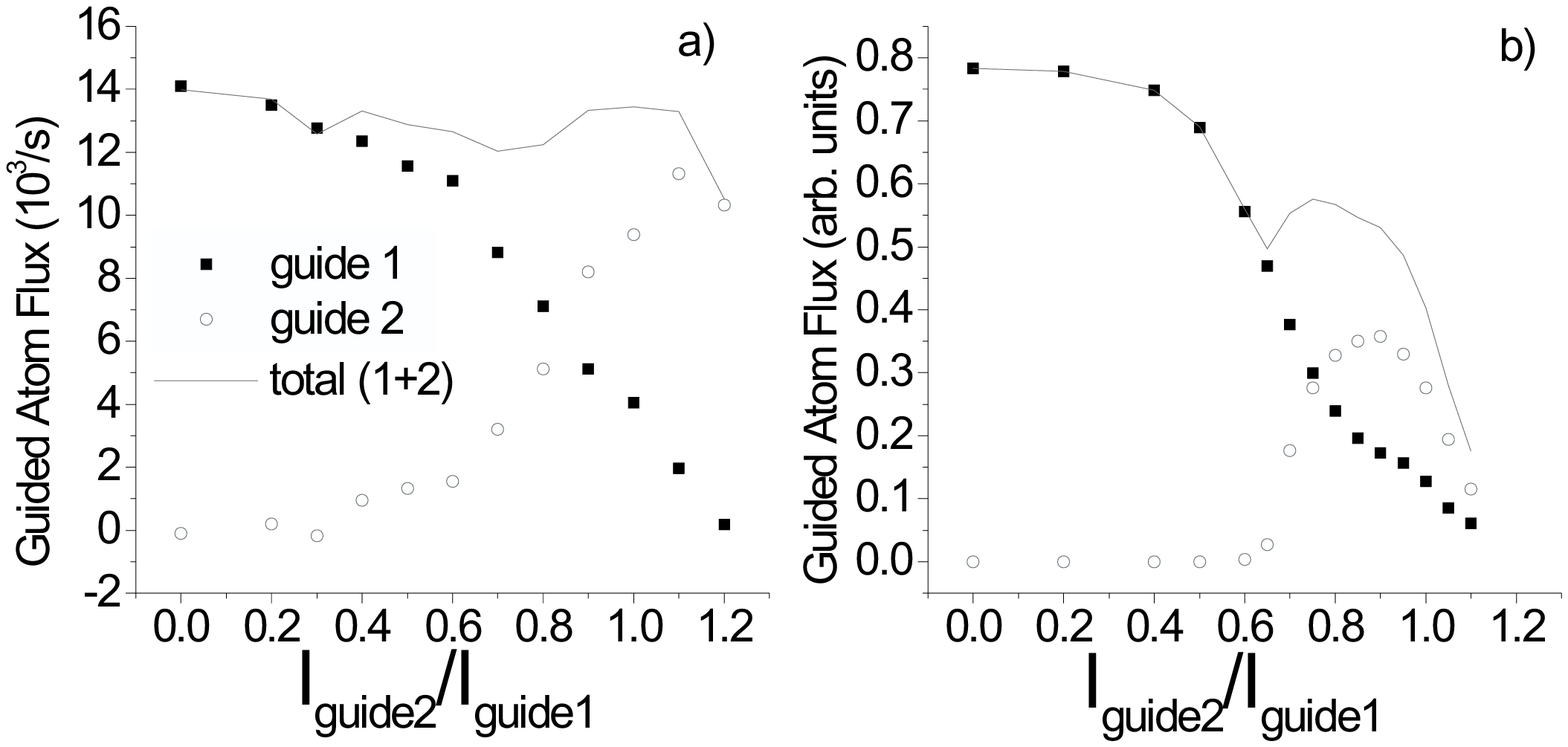}

\caption{Flux versus wire-current ratio. Figure 3(a) shows the
experimental data and figure 3(b) a simulation of our
beamsplitter. With no current in wire 2 all atoms coupled into
guide 1 exit the same guide 1. As the current in wire 2 is
increased the beamsplitter is turned on and atoms are transferred
to guide 2.  For the experimental (simulation) data the
transverse-bias field, the longitudinal-bias field, and $I_{guide
1}$ are held constant at values 86 G (100 G), 14 G (14 G), and 5.0
A (5.0 A) respectively. The simulation shows good qualitative
agreement with our experimental data. \label{fig3}}
\end{figure}

\pagebreak

\begin{figure}
\epsfxsize=3 truein \epsfbox{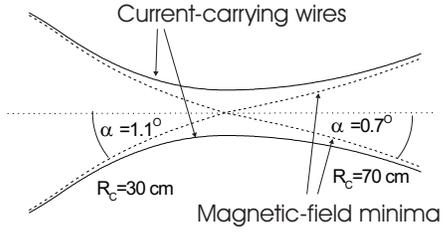}

\caption{Magnetic-field-minima crossing.  For our curvature
($R_{c}$) and spacing the magnetic field minima cross in the
center region with an angle $\alpha$ between the two trajectories.
Atoms coupled into one guide are more likely to be transferred
into the other guide when both wires carry equal currents, because
of the tendency to shoot straight across the intersection.
\label{fig4}}
\end{figure}

\end{document}